# Modelling Dead Rocking In Online Multiplayer Games

*Abstract*— This is the web based application, here we analysis the network traffic which occurs when the player plays an online game. Here we are going to trace the current position of the player to rectify the traffic while playing the game. There are different types of measures for different applications, those can be normalized and compared with one another but my application can resolve the inconsistency by knowing the positions quickly and focus on quality of network (QON) which affects a player to leave the game in middle because of poor quality of service (QOS). The existing model leads to leave the game because of network loss from both sides. The proposed model can resolve this problem by applying the replacement of TCP along with the dejitter buffer (DB) it can reduce the network loss and by applying the DEAD RECKONING (DR) vector we can recover the network access because it can view the current position of the player through this we can rectify why the player leaving the game and checks the network conditions and try to reactive the game by using mobile devices automatically or else they receive the message according to the network conditions.

## 1. INTRODUCTION

Networking plays an essential role in multiplayer computer games. In this Paper, we audit the techniques developed for improving the networking in distributed interactive real-time applications. We also discuss the resource management, consistency and responsiveness, and networking on the application level. Of the various research areas related to online games, assessing the impact of network conditions on user experience is one of the most popular topics. For instance, if we can be sure that players are less tolerant of large delay variations than high latency, then providing a smoothing buffer at the client side, which introduces additional latency but smoothes the pace of game play, would be a plus, as it still improves the overall gaming experience from the user's perspective. Currently, there is no standard way to objectively quantify the satisfaction that players derive from gaming. Hence, the effect of network quality is often evaluated in terms of the users' performance in a specific context such as the number of kills in shooting games, the time taken to complete each lap in racing games, or the capital accumulated in strategy games.

## 2. MODEL

As online games, especially MMOG (Massive Multiplayer Online Games), grow popular, the share of game traffic on the Internet has become increasingly significant. Reported in a backbone traffic analysis, about 3%–4% of the traffic is attributed to 6 popular games. Given the significant share of game traffic and the dissimilar nature of games from dominant Internet applications such as the World Wide Web, peer-to-peer filing sharing, and streaming, a better understanding of the game traffic is vital. In a distributed multi-player game, players are normally geographically distributed. In such games, the players are part of the game and in addition they may control entities that make up the game. During the course of the game, the players and the entities move within the game space. A technique referred to as Dead Reckoning (DR) is commonly used to exchange information about player/entity movement among the players. A player sends information about her movement as well as the movement of the entities she controls to the other players using a Dead Reckoning (DR) vector. A DR vector typically contains information about the current position of the player/entity in terms of x, y and z coordinates (at the time the DR vector was sent) as well as the trajectory of the entity in terms of the velocity component in each of the dimensions. Each of the players that form part of the game receives such DR vectors from each other and renders the other players/entities on the local console until a new DR vector is received for that player/entity. Thus, the DR vector is a model of expected future behavior, and is used to "predict" the position of the player/entity until an updated DR vector is received. Normally, a new DR vector is sent whenever the path of the player/entity at the sender deviates from the path corresponding to the previous DR vector (say, in terms of distance in the x, y, z plane) by some amount specified by a threshold. In a peer-to-peer game, players send DR vectors directly to each other; in a client-server game, these DR vectors may be forwarded through a game server. The idea of DR vector is used because it is almost impossible for players/entities to exchange their current positions continuously or at some very small time unit.

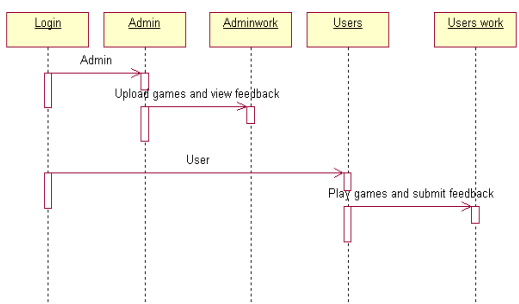

Fig 1 Sequence Diagram

As online games, especially MMOG (Massive Multiplayer Online Games), grow popular, the share of game traffic on the Internet has become increasingly significant. Reported in a backbone traffic analysis, about 3%–4% of the traffic is attributed to 6 popular games. Given the significant share of game traffic and the dissimilar nature of games from dominant Internet applications such as the World Wide Web, peer-to-peer filing sharing, and streaming, a better understanding of the game traffic is vital. We aim particularly at MMORPG (Massive Multiplayer Online Role Playing Games) for two reasons. First of all, MMORPG is the most prominent game genre in Asia. In Taiwan, Gamania1, operator of the popular game – Lineage 2, owns more than 4,000 Mbps dedicated links3 for game traffic. According to Game Flier4, the top game, Ragnarok Online, has claimed a record of 370,000 players online simultaneously. This is about 1.5% of the population in Taiwan. Further more, most MMORPG in Asia exchange messages using TCP. Although TCP is not designed for real-time communication, it is not clear as yet whether TCP is not suitable for the MMORPG traffic transmission. To the best of our knowledge, this work is the first analyzing and characterizing the MMORPG traffic. We trace Shen-Zhou Online [1], a mid-sized MMORPG, and analyze a 1,356-million-packet trace. Our analysis focuses on the packet size, bandwidth usage, packet inter arrival within a connection, and the arrival process of the aggregated traffic. The major findings are summarized as follows:

• Most of the packets are *small*. 98% packets sent by the game clients are smaller than 71 bytes. This suggests that the overhead of packet headers and TCP acknowledgments is high relative to other Internet applications. In total client traffic, headers occupy 73% bytes, and TCP acknowledgement packets take up 30%.

• The average bandwidth requirement per client is about 7 Kbps, which is *much lower* than the 40 Kbps average observed from Counter-Strike. We believe the lower bandwidth requirement is due to the relatively slow motion or action pace in MMORPG.

• The traffic of either direction exhibits short-term positive auto-correlations within connections. In the client traffic (traffic going from the client to the server), the positive auto-correlations is due to the *temporal locality in user actions*. In the meantime, the same effect in server traffic is due to the *spatial locality in the number of nearby characters*. The spatial locality shows up in terms of temporal locality in the traffic as the characters move continuously on the map.

• Positive auto-correlations still exist in aggregated client traffic. We consider it is owing to the *global events* in games, which cause the "flash crowds" effect. Furthermore, the aggregated traffic of either direction exhibit strong *periodicity*. It implies the game processing for all clients are synchronized. The remainder of this paper discuss these issues in more detail describes related works and the game Shen-Zhou *Online*. We show the trace measurement methodology and trace summary in we present a detailed analysis on the game traffic from various aspects. With the launches of Microsoft's Xbox on-line game network and Sony's Play station 2 n-line game network, along with the development of emerging massively multi-player on-line games that allow for thousands of simultaneous players, it is clear that gaming traffic will soon grow to scales far beyond what is being observed today. While there has been a lot of work characterizing aggregate web traffic behavior as well as web client behavior, there has been very little work in doing the same for games. Because of the tremendous increase in game traffic, it is important to understand the traffic demands that this emerging application imparts. While not indicative of all on-line games, the class of games known as "first-person shooters" has clearly dominated much of the observed gaming traffic. In this paper, we analyze the session behavior of game clients playing on a popular FPS game server and develop a source model for their ON times. We note that previous work has shown that when such clients are active, they impart almost a constant bit-rate load on the network. Describes a detailed game server trace we performed in order to analyze the session behavior of players in on-line games. Section 3 analyzes the session times of players in the trace and develops a source model that describes session times of players in the trace. Concludes with a discussion of the implications that these results have on traffic characterization of on-line games.[2]

In recent years, the declining costs of increasingly powerful personal computers has increased their acquisition rates and created a growing user base for computer games. The increase in residential broadband Internet connections with high capacities and low latencies have encouraged more and more game developers to incorporate multi-player features into their products. An understanding of how network related issues, such as latency and packet loss, affect the usability of games can be of great use to the companies that make these games, network software and equipment manufacturers, Internet Service Providers (ISPs), and the research community at large. In particular, once established latency requirements and any associated trade-offs are known, ISPs can establish tariffs based on customers' indicated maximum delays, requested Quality of Service (QoS), and the ISP's ability to meet these demands. Moreover, experimental study of network games can provide the data required for accurate simulations, a typical tool for evaluating network research, as well as insight for network architectures and designs that more effectively accommodate network game traffic turbulence. While there has been research qualitatively characterizing the effects of latency for car racing, custom games, and real-time strategy games as well as a general awareness of latency issues work on the effects of latency in popular First Person Shooter (FPS) games has not quantified its impact on player performance.[3] In concentrating on the effects of latency on FPS games, the possibility that packet loss may be the bottleneck in performance for some network conditions may be overlooked. To the best of our knowledge, there have been no systematic studies of packet loss on user performance in FPS games. The study of loss on network games is increasingly important as wireless channels, more prone to packet loss than traditional wire-line environment, become widely adopted. In general, the most popular FPS games have descended from two game lineages, using either a Quake or UN real based game engine. As previous research has concentrated on FPS games derived from Quake, we used Epic Game's award winning1 Unreal Tournament 20032 (UT2003) in our experiments. UT2003 is very popular, with approximately 1700 servers and 4400 players online at any given time. First, we deduced typical real world values of packet loss and latency experienced on the Internet by monitoring operational UT 2003 game servers. We then used these values as guidelines for induced loss and latency values in a controlled emulated environment we designed, allowing us to monitor UT2003 at the network, application and user levels. We divided user interaction in UT2003 into fundamental FPS interaction components in order to isolate particular facets of game-play. These interaction components include movement, precision shooting, general shooting, and moving and shooting simultaneously. We designed experiments with game maps that allowed us to isolate each game component. Using our test bed, we ran numerous user studies during which we systematically changed the loss and latency and measured the impact on performance. We find that the levels of packet loss and latency typically encountered on the Internet, while sometimes unpleasant, do not drastically impact user performance in UT2003. Loss, in particular, goes unnoticed and does not measurably affect user interaction. Latencies as low as 100 ms, on the other hand, can significantly degrade performance for shooting with precision weapons both in terms of accuracy and game responsiveness. Although the effects of latency on user performance in full UT2003 games with all components is less noticeable, there is still a clear user performance degradation trend as latency increases.[4]

### 3. Experimental results

In a distributed multi-player game, players are normally geographically distributed. In such games, the players are part of the game and in addition they may control entities that make up the game. During the course of the game, the players and the entities move within the game space. A technique referred to as *Dead Reckoning* (DR) is commonly used to exchange information about player/entity movement among the players .A player sends information about her movement as well as the movement of the entities she controls to the other players using a Dead Reckoning (DR) vector. A DR vector typically contains information about the current position of the player/entity in terms of x, y and z coordinates (at the time the DR vector was sent) as well as the trajectory of the entity in terms of the velocity component in each of the dimensions. Each of the players that form part of the game receives such DR vectors from each other and renders the other players/entities on the local console until a new DR vector is received for that player/entity. Thus, the DR vector is a model of expected future behavior, and is used to "predict" the position of the player/entity until an updated DR vector is received. Normally, a new DR vector is sent whenever the path of the player/entity at the sender deviates from the path corresponding to the previous DR vector (say, in terms of distance in the x, y, z plane) by some amount specified by a threshold. In a peer-to-peer game, players send DR vectors directly to each other; in a client-server game, these DR vectors may be forwarded through a game server. The idea of DR vector is used because it is almost impossible for players/entities to exchange their current positions continuously or at some very small time unit.

### 4. Advantages

We analyze the player departure patterns in ShenZhou Online and their relationship to network quality. We find that both network latency and network loss have a significant influence on players' decisions to leave a game prematurely. As the number of user's increases in an online game the efficiency of the service will not be degraded. Scalability of the services doesn't depend up on the number of user in online.

### 5. Conclusion

To understand the relationship between network quality and players' departure patterns, we analyzed a 1,356-millionpacket trace from a commercial MMORPG called ShenZhou Online. Our results indicate that both network delay and network loss significantly affect a player's decision to leave a game prematurely, i.e., the player quits a few minutes after joining a game. We show that it is feasible to predict whether players will quit prematurely based on the network conditions they experience. The proposed model can determine the relative impact of different types of network impairment and try to reactive the game by using mobile devices automatically or else they receive the message according to the network conditions.